# Refinement in the Function-Behaviour-Structure Framework (version 2)

*Bob Diertens*

section Theory of Computer Science, Faculty of Science, University of Amsterdam

### ABSTRACT

We introduce refinement in the function-behaviour-structure framework for design, as described by John Gero, in order to deal with complexity. We do this by connecting the frameworks for the design of two models, one the refinement of the other. The result is a framework for the design of an object that supports levels of abstraction in the design. This framework can easily be extended for the design of an object on more than two levels of abstraction.

## 1. Introduction

In software engineering, dealing with complexity is a major issue and it is the ground for many software development methodologies. Most of these methodologies however do not take into account the nature and process of design. Each methodology has its success stories, but one can seldom relate it to a more abstract framework for design. A well-known method for dealing with complexity in other engineering disciplines is modelling. By making a model one can leave out some detail and concentrate on the bigger picture. Even a model can be too complicated, in which case one can make a model for the model. This results in a design consisting of several levels, each higher level abstracting from details on the lower levels.

This is picked up in software engineering as well and resulted in model based software development and similar approaches. Despite this, the problem with software engineering remains that the design is largely focused on a too low level of abstraction. This is caused by the fact that software is build cheaply, and can be done over and over again. This makes it possible to test on the lowest level and often results in a race to the lowest level to start testing early in the design process. A step up to a higher level of abstraction to repair design flaws is easily replaced by fixing the design on the lowest level. In that process, the higher level design is discarded and complexity is taken into the lower levels instead of dealing with it on the higher levels of design.

In our view it is better to incorporate methodologies that follow the nature and process of design on several levels of abstraction. An important factor in this is the knowledge of what design really is. John Gero has described a general framework for design [1] that is based on function, behaviour, and structure of the object to be designed. This framework, however, omits levels of abstraction. Although one can argue that through the reformulation processes of this framework it is possible to have levels of abstraction in the design, the levels of abstraction then are only implicit in the design process. For a thorough understanding and execution of the design process it would be better to make the levels of abstraction explicit in the design process.

In section 2 we give an overview of the function-behaviour-structure framework for design. We introduce refinement in this framework in section 3 in order to support levels of abstraction in the design process explicitly.

## 2. The Function-Behaviour-Structure Framework

In [1] Gero describes a framework for design that has sufficient power to capture the nature of the concepts that support design processes. This framework, that involves the relation between function, behaviour, and structure of a design, can be applied to any engineering discipline. Together with Kannengiesser, Gero describes the framework in [2] in relation with the environment in which designing takes place, accounting for the dynamic character of the context. We give an overview of the elements and processes that form the function-behaviour-structure (FBS) framework.

The FBS framework elements has the following elements.

| | |
|---|---|
| function ($F$) | The set of functions expressing the requirements and objectives that must be realized by the object. |
| structure ($S$) | Describes the components of the object and their relationships. |
| expected behaviour ($B_e$) | The set of expected behaviours to fulfill the function $F$. |
| structure behaviour ($B_s$) | The set of behaviours the structure $S$ exhibits. |
| description ($D$) | The description of the design, giving all the information to build the object, and what more there is to know about the design. |

These elements are connected in the framework by processes (Figure 1).

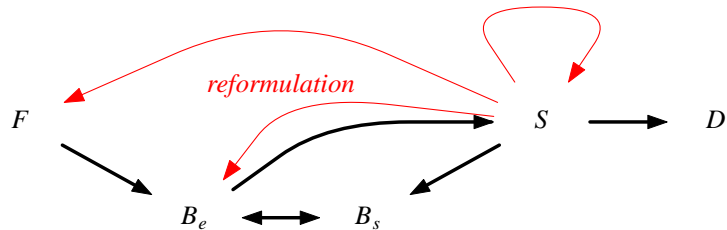

**Figure 1.** The FBS framework

An outline of the process of the FBS framework is given below.

| | |
|---|---|
| formulation ($F \rightarrow B_e$) | Transforming the function $F$ into behaviour that is expected from the object. |
| synthesis ($B_e \rightarrow S$) | Transforming the expected behaviour into a solution intended to exhibit this behaviour. |
| analysis ($S \rightarrow B_s$) | Deriving of the actual behaviour from the synthesized structure. |
| evaluation ($B_e \leftrightarrow B_s$) | Comparing the behaviour derived from the structure with the expected behaviour. |
| documentation ($S \rightarrow D$) | Producing the design description for the constructing or manufacturing of the object. |

In addition the framework contains reformulation processes that are carried out, based on the outcome of the evaluation of behaviours.

structure reformulation ($S \rightarrow S$)
: Changing of the structure in order to obtain a behaviour that is more in line with the expected behaviour.

behaviour reformulation ($S \rightarrow B_e$)
: Adjusting of the expected behaviour that fits the required function and is more in line with the behaviour of the structure.

function reformulation ($S \rightarrow F$)
: Changing of the function due to a better insight in the problem.

## 3. Refinement of the FBS framework

To capture refinement in design we combine FBS frameworks into a framework ($R-FBS$) that shows different levels of abstraction. We consider the design of two models, $M$ and $M'$, for a particular system. The model $M'$ is supposed to be a faithful implementation of the model $M$. This has as a consequence that the two models both represent the system but on different levels of abstraction. Both models have their own design process, $FBS$ and $FBS'$, each of which can be described by the function-behaviour-structure framework for design.

Because the model $M'$ on the lower level of abstraction is an implementation of the model $M$ on the higher level of abstraction, the description $D$ for model $M$ contains information for the design of model $M'$. Because the model $M'$ is a refinement of the model $M$, the structure $S'$ in the design process $FBS'$ is a refinement of the structure $S$ in the design process $FBS$. The relations described above between the elements in the design processes $FBS$ and $FBS'$ for the two models is shown in Figure 2.

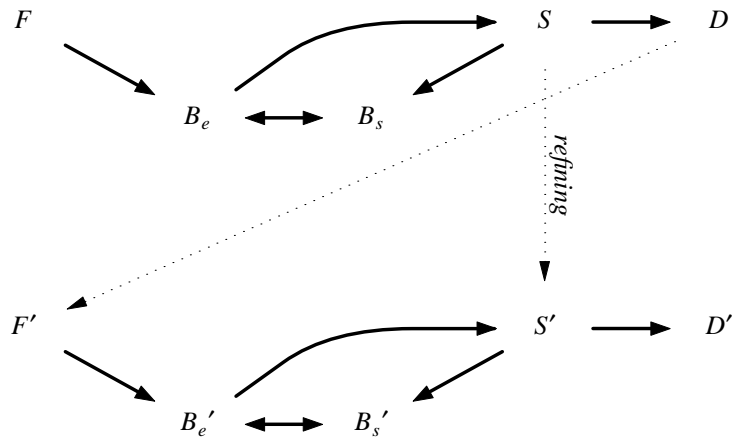

**Figure 2.** Design framework for two models

We like to integrate the two design processes, so that the processes that play a role in the refinement, evaluation of the refinement, and reformulation of the refinement become clear. In the following sections we describe how these design processes can be interconnected and what the consequences are for the overall design process.

*3.1 Refinement*

Each element of $FBS'$ can be considered a refinement of the associated element of $FBS$. In the following we describe how the refinement processes between the two frameworks take place.

> Functionality refinement ($\{D, F\} \rightarrow F'$)
> As the structure $S$ consists of the components and their interaction for model $M$, the description $D$ describes the functionality of these components and the interactions. It is this functionality together with functionality $F$ that makes up the functionality $F'$ for model $M'$. This refinement of functionality is indicated by 1 in Figure 3.

> Expected behaviour refinement ($\{B_e, F'\} \rightarrow B_e'$)
> The expected behaviour $B_e'$ can be obtained by refining the expected behaviour $B_e$ with refinements extracted from $F'$. This is indicated by 2 in Figure 3.

> Structure refinement ($\{S, B_e'\} \rightarrow S'$)
> In a FBS framework the structure is synthesized from the expected behaviour of the model. We cannot obtain the structure $S'$ from $B_e'$ in this way, since $S'$ must be a refinement of $S$. We have to synthesize $S'$ from $S$ and use refinements that are based on $B_e'$. This is indicated by 3 in Figure 3.

Documentation refinement ($\{D, S'\} \rightarrow D'$)

The description $D'$ is the addition of the description $D$ and the description of the refinement processes. This is indicated by 4 in Figure 3.

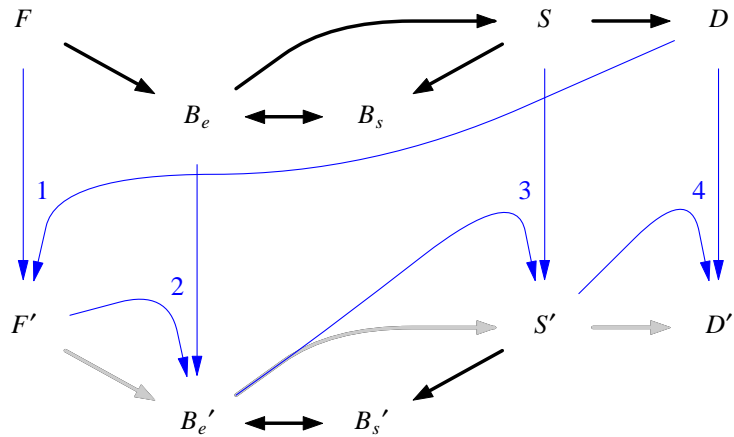

**Figure 3.** Refinement processes in the design framework

## 3.2 Behaviour Evaluation

Besides the behaviour evalution on both levels, we have to check if $M'$ is a true implementation of $M$. We can do this by comparing a refined $B_s$ with $B_s'$. But that leaves us with the problem how to refine $B_s$. We can however do the reverse. Instead of refining $B_s$ we can abstract $B_s'$ and compare it with $Bs$ (Figure 4). If the behaviours do not match, the refinement process is wrong and has to be adjusted. The adjustment can be done through reformulation processes described in the next section.

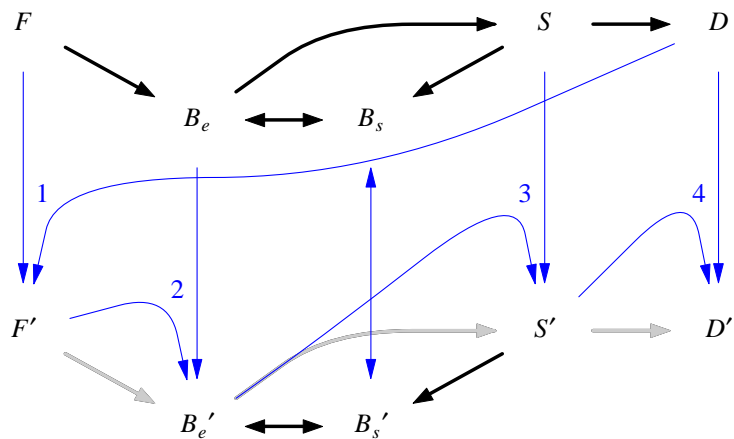

**Figure 4.** Refinement evaluation process in the design framework

*3.3 Reformulation*

The design frameworks for the models $M$ and $M'$ contain reformulation processes. For the model $M$ these are the standard processes described earlier, but for the model $M'$ the reformulations have to be such that the elements stay within the refinements of their corresponding abstract elements.

The situation can occur that a reformulation of one of the elements for model $M'$ is necessary and that because of this reformulation it no longer conforms with the corresponding abstract element for model $M$. In that case, the current design must be rejected and a reformulation of the corresponding element for model $M$ have to take place. All the reformulation processes are shown in Figure 5.

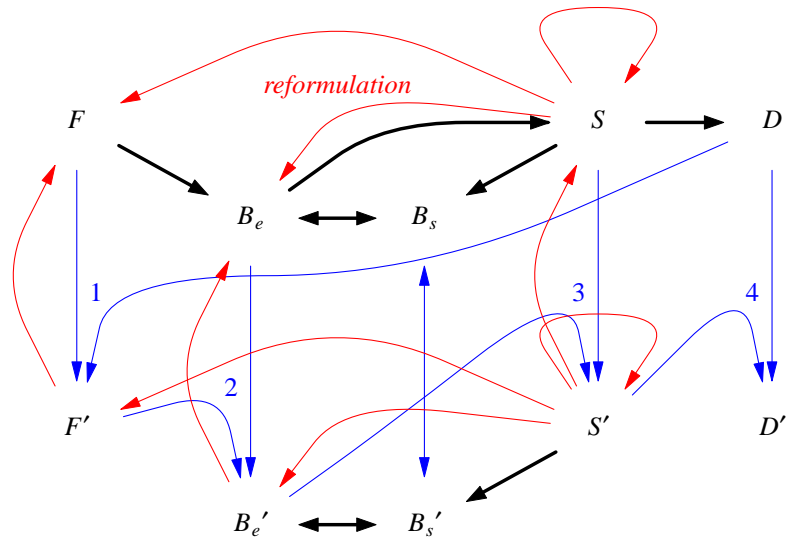

**Figure 5.** Reformulation processes in the design framework

## 4. Conclusions

We introduced refinement in the FBS framework by connecting frameworks for the design of two models, one model the refinement of the other. The resulting framework ($R-FBS$) can be used for the design of an object on two levels of abstraction. In a similar way, $R-FBS$ can easily be further extended to a framework for the design of an object on more than two levels of abstraction. We can turn $R-FBS$ into the original framework by considering the refinement processes as reformulations and abstract from the details of the refinement processes. In $R-FBS$ the levels of abstraction in the design are made explicit. This $R-FBS$ can be used to develop several models that are related to each other through refinement and abstraction.

## Acknowledgements

Many thanks to Alban ponse for his proofreading and feedback.

## References

[1]  J.S. Gero, ''Design Prototypes: A Knowledge Representation Scheme for Design,'' *AI Magazine*, vol. 11, no. 4, pp. 26-36, 1990.

[2]  J.S. Gero and N. Kannengiesser, ''The Situated Function-Behavior-Structure Framework,'' *Design Studies*, vol. 25, no. 4, pp. 373-391, 2004.